\documentclass{ieeeaccess}

\usepackage{etoolbox}
\makeatletter
\def\ps@headings{\let\@oddhead\@empty\let\@evenhead\@empty
  \def\@oddfoot{\hfil\footerpagefont\thepage\hfil}\let\@evenfoot\@oddfoot}
\let\ps@titlepage\ps@headings
\patchcmd{\@maketitle}{\ifaccesseditors\else{\vss\historyfont\@history\vss}\vspace*{2mm}\par\fi}{}{}{\PackageWarning{arxiv}{history patch failed}}
\patchcmd{\@maketitle}{{\vss\doifont Digital Object Identifier\space\@doi\vss}\vspace*{7.4mm}\par}{}{}{\PackageWarning{arxiv}{doi patch failed}}
\makeatother
\def\EOD{\global\eoddefinedtrue}% keep the class happy, print no end mark
\usepackage{cite}
\usepackage{amsmath,amsfonts,amssymb}
\usepackage{algorithm}
\usepackage{algorithmic}
\usepackage{array}
\usepackage{booktabs}
\usepackage{textcomp}
\usepackage{stfloats}
\usepackage{url}
\usepackage{verbatim}
\usepackage{graphicx}
\newdimen\xfigwd
\begin{document}

\history{Date of publication: to be completed by IEEE.}
\doi{DOI to be assigned by IEEE}

\title{Multi-Label Classification with Constrained Search for Inverse Design of a 16x16 Reconfigurable Pixel Antenna}

\author{\uppercase{Nadeem Rather}\authorrefmark{1},
\uppercase{Holger Claussen}\authorrefmark{2},
\uppercase{Lester Ho}\authorrefmark{1}}

\address[1]{Tyndall National Institute, Cork, Ireland (e-mail: nadeem.rather@tyndall.ie; lester.ho@tyndall.ie)}
\address[2]{Tyndall National Institute, Cork, Ireland; University College Cork, Cork, Ireland; and Trinity College Dublin, Dublin, Ireland (e-mail: holger.claussen@tyndall.ie)}

\corresp{Corresponding author: Nadeem Rather (e-mail: nadeem.rather@tyndall.ie).}

\begin{abstract}
In this paper, an inverse design method is presented for a $16 \times 16$
reconfigurable pixel antenna with 480 binary switches, giving $2^{480}$ possible
patterns, far beyond what full-wave simulation can explore. Given a target frequency between 1 and 6~GHz, the method searches for a switch configuration whose simulated resonance lies close to the target. A forward surrogate model is
trained on 22{,}152 full-wave simulated patterns to predict where a given switch
pattern resonates. The forward task is formulated as multi-label classification over
100 frequency bins rather than reflection-coefficient regression, because a
curve-fitting objective is dominated by the flat baseline around the narrow resonances
of this structure. The inverse design is a simulated annealing search over the switch state, with
a connectivity constraint maintained at every step so that every pattern generated is
a valid antenna. The proposed methodology combines resonance-focused surrogate modelling with connectivity-constrained discrete optimization, enabling candidate generation without full-wave solver calls during the search. The pipeline is verified in CST Microwave Studio by simulating 220 predicted
candidates, ten for each of 22 target frequencies. For each request, the best of the ten candidates reached a median frequency error of
23~MHz across the 22 targets, compared with 35~MHz for a solver-guided search using 200 simulations for each target. The proposed methodology therefore enables automatic switch configuration
to achieve a requested resonant frequency, eliminating solver calls during the search
and manual redesign of the antenna geometry.
\end{abstract}


\maketitle
\vspace{-3mm}
% =====================================================================
\section{Introduction}
\label{sec:intro}
\PARstart{R}{econfigurable} planar antennas have become central to modern wireless communication systems, where a single low-cost structure is expected to adapt its behaviour on demand \cite{christodoulou2012,costantine2015}. This adaptability is achieved by tuning elements such as switches, PIN diodes, or varactors, which allow the operating frequency, radiation pattern, or polarization of the antenna to be changed after fabrication \cite{hum2014,haupt1995}. In one widely used approach these switches are placed between a grid of small metallic elements, known as pixels, so that the connections between pixels are changed and the radiating structure is reshaped when the switches are turned on or off. 
\begin{figure}[!t]
\centering
\includegraphics[width=\columnwidth]{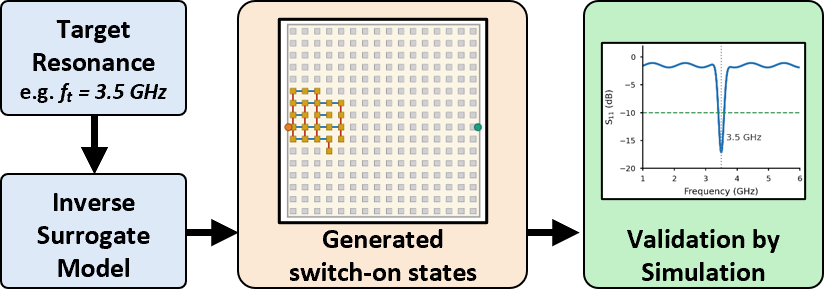}
\caption{Overview of the proposed inverse design method: a target resonant frequency is specified, the inverse search produces a switch pattern, and the design is verified in CST Microwave Studio.}
\label{fig:overview}
\vspace{-4mm}
\end{figure}
This class of structures, known as pixel antennas, is valued for its high reconfigurability, low cost, and compact form factor \cite{jing2022,costantine2015}. By setting the state of every pixel connection, the current distribution, impedance match, and resonant behaviour of the antenna can be engineered, which makes pixel antennas a flexible platform for programmable wireless systems \cite{direnzo2020,liaskos2018}. The central problem in using such a board is inverse design. The user
specifies a desired electromagnetic response, for example a resonant
frequency, and the design task is to find a switch pattern that produces
it. With $N$ binary bridges the design space contains $2^N$ patterns. For
the antenna considered in this paper $N=480$, so the design space has
approximately $3.1 \times 10^{144}$ elements. Direct enumeration is not
possible. The proposed method is shown in Fig.~\ref{fig:overview}. A target frequency is
specified, a switch pattern is found by a search guided by a trained surrogate model,
and the result is verified by full-wave simulation.

Two families of methods have been used for related problems. In topology optimization the design is treated as a continuous density field and gradients of the electromagnetic objective with respect to that field are used \cite{bendsoe1988,christiansen2021}. This approach is well established in mechanics and has been extended to nanophotonics and antenna design \cite{molesky2018}. In genetic algorithms and other evolutionary methods the design is treated as a discrete vector and explored by mutation and selection \cite{haupt1995,deb2002}. Both families share a structural difficulty: every candidate evaluation requires a full-wave electromagnetic simulation that takes several minutes on a workstation. The number of evaluations a typical budget allows is therefore in the hundreds or low thousands, and the resulting designs are often small or symmetric to keep the search tractable. Machine learning has been used to remove this evaluation bottleneck. A surrogate model is trained on a moderate number of full-wave simulations and is then used in place of the solver during the search \cite{molesky2018,wiecha2021,ma2021,salucci2018,elmisilmani2023}. Once trained, a candidate design is evaluated by the surrogate in milliseconds. The forward surrogate can either be inverted directly by gradient-based optimization on the model input \cite{peurifoy2018,liu2018} or used as a scoring function inside a discrete search such as a genetic algorithm or simulated annealing. The first approach is fast, but a smooth input space is required; the second is slower, but applicable to non-smooth or constraint-heavy designs.
For pixel antennas, where the design variables are binary, the evaluation cost has
been reduced in two ways. In the first, every switch location is modelled as an
internal port, so that one multiport simulation allows any switch state to be
evaluated by circuit analysis. This model has been combined with genetic
algorithms~\cite{song2014} and with successive exhaustive Boolean optimization,
demonstrated on a $5 \times 5$ pixel antenna with 44 possible
connections~\cite{shen2017}. Every evaluation still solves a network with one port
per switch at each frequency point, so its cost rises steeply with the size of the
board. In the second, an ensemble-learning surrogate guides an evolutionary search
for digitally coded antennas of more than a thousand pixels~\cite{wu2025}. Each run
produces one fixed antenna for one specification, so a reconfigurable board would
need a new optimisation for every requested state.

Within antenna design specifically, machine learning has been applied to the inverse problem in several forms. Tandem networks have been used to design multiband antennas by training a forward model and an inverse model together \cite{gupta2023}, an inverse artificial neural network has been used to map a requested response back to geometric parameters under multiple objectives \cite{xiao2021}, and deep networks have been used to generate microstrip antenna geometries from a specified response \cite{chen2025}. In these methods, the geometry is assumed to be defined by a small number of continuous parameters, such as lengths, widths, and slot positions, which enables direct regression of these parameters using an inverse network. That assumption does not hold for the structure considered here, where the design is a 480-element binary vector with no continuous parametrisation and where a single bit change can move a resonance by hundreds of megahertz. \par

Direct regression of the reflection-coefficient curve from the switch pattern has been the standard formulation in much of this literature, including pixelated antennas trained with a loss weighted toward the resonance~\cite{rather2026}, but it is a poor match for the narrow resonances of the board considered here.  \par
\vspace{2mm}
The geometry of a pixel antenna cannot be described by a few shape parameters, which is what makes its response hard to model in the first place \cite{jacobs2021}. More importantly, the resonances of these structures are narrowband. A deep dip in the reflection coefficient occupies only a small part of the frequency axis, while the response stays close to its baseline everywhere else. Two effects follow from this. First, neural networks generally approximate smooth, slowly varying functions more readily than sharp local features, making narrow dips particularly difficult to reproduce accurately \cite{rahaman2019}. Second, when the quantity of interest is rare relative to the baseline, a root-mean-square error is dominated by many baseline points, and standard error metrics are reduced more by fitting the common case than the rare one \cite{ribeiro2020}. A model can therefore drive the training error down by predicting a nearly flat curve that follows the baseline and misses the resonances entirely, which is the behaviour observed when the full curve is regressed on the data used here. This makes root-mean-square error an unreliable training objective for this problem, motivating a formulation in which the presence of a resonance, rather than the exact curve value, becomes the quantity the model predicts.

In this paper, the inverse design problem is addressed for a specific $16 \times 16$ reconfigurable pixel antenna with 480 binary switches. The approach has three components:
\vspace{2mm}
\begin{enumerate}
\item The forward surrogate task is reformulated from scattering-parameter regression to multi-label classification over a discretized frequency axis. Each pattern is labelled by a 100-bin vector of Gaussian soft labels wide enough to span several bins, so that a narrow resonance still produces a usable training signal.
\vspace{3mm}
\item The inverse search is a simulated annealing chain over the 480-bit switch vector. The connectivity constraint is maintained at every step, so every state generated during the search corresponds to a valid antenna. By applying a soft penalty on the number of active patches, the search is constrained to the region of input space where the forward surrogate has good training coverage.
\vspace{3mm}
\item The complete pipeline is verified by full-wave simulation of 220 generated candidates spanning 22 target frequencies, rather than by reporting forward-model self-evaluation only.
\end{enumerate}

Together, these components address both limitations of the earlier pixel-antenna
methods. On a board with 480 switches, more than ten times the connections
in~\cite{shen2017}, a candidate state is evaluated in about 3~ms through a single
forward pass of the surrogate rather than a network solve over every switch. The model
is trained once, and each new frequency request is then answered by a 2.2-minute
search with no solver calls.

The remainder of this paper is organized as follows: The reconfigurable pixel antenna and its switching topology are described in Section~\ref{sec:hardware}. The pattern generator, simulation protocol, and dataset statistics are detailed in Section~\ref{sec:dataset}. The forward surrogate model and its training procedure are outlined in Section~\ref{sec:forward}, followed by the inverse search strategy in Section~\ref{sec:inverse}. In Section~\ref{sec:results}, the performance metrics of the forward model, full-wave verification of the complete pipeline, and a comparison against a solver-guided search are presented. \par Furthermore, an extension of the method to two-port filter operation and the limitations of the present work are addressed in Section~\ref{sec:discussion}, followed by the conclusion in Section~\ref{sec:conclusion}. 

% =====================================================================
\section{Reconfigurable Pixel Antenna}
\label{sec:hardware}
\begin{figure}[!t]
\centering
\includegraphics[width=\columnwidth]{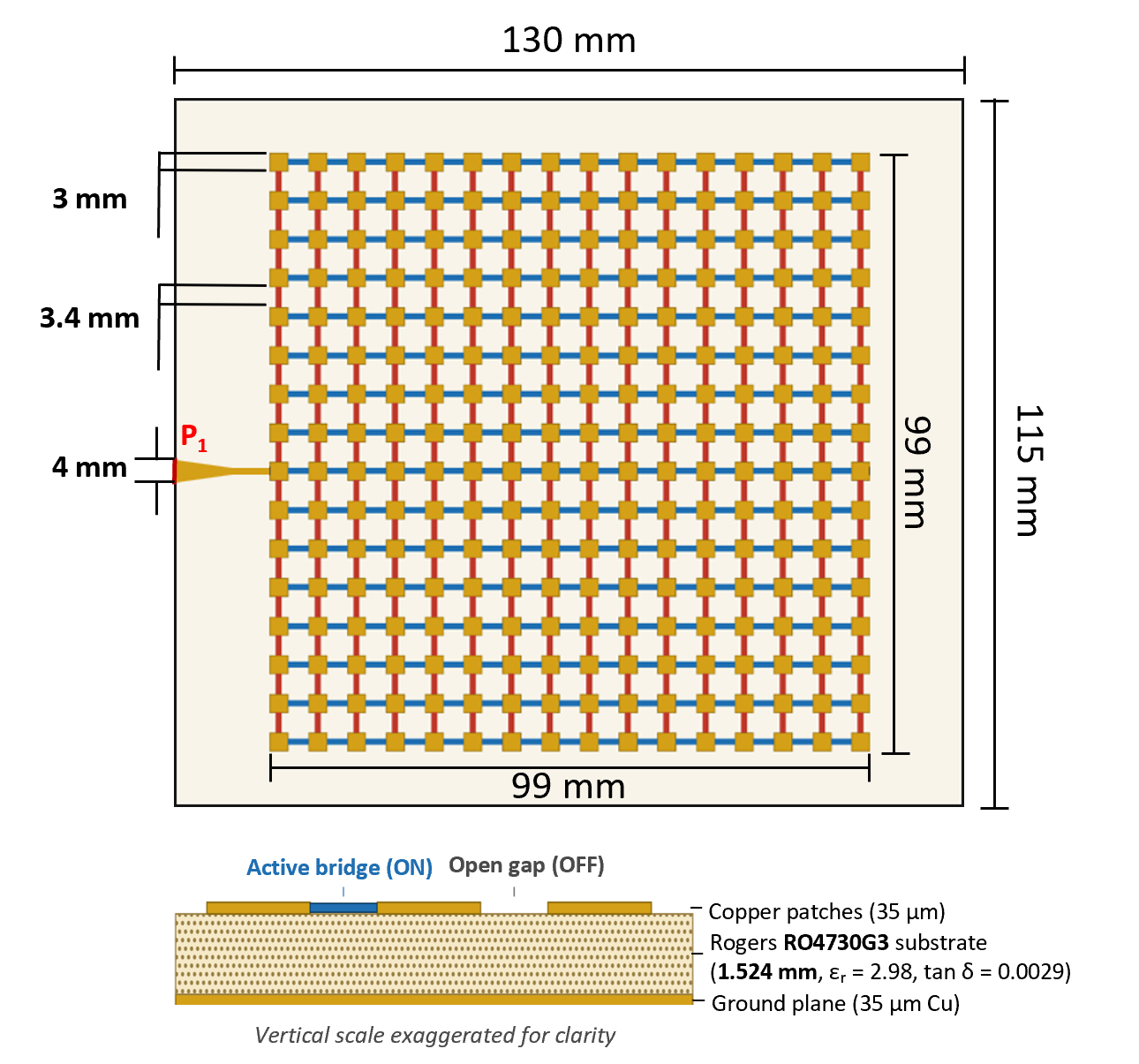}
\caption{Top-down and side view of the $16 \times 16$ reconfigurable pixel antenna. A representative pattern is shown with shaded patches and active bridges. Port P1 is fixed at column~1, row~8 in 1-based indexing.}
\label{fig:hardware}
\end{figure}

\vspace{2mm}
The structure is a $16 \times 16$ lattice of square copper patches, each $3 \times 3$~mm and 35~$\mu$m thick, modelled on a Rogers RO4730G3 substrate of thickness 1.524~mm, relative permittivity $\varepsilon_r = 2.98$, and loss tangent $\tan\delta = 0.0029$. Adjacent patches are separated by a 3.4~mm gap, and the lattice spans $99 \times 99$~mm edge to edge. A continuous copper ground plane of dimensions $130 \times 115$~mm underlies the structure. The board dimensions match the band of interest. To first order, a conducting path
over the ground plane resonates when its length is close to half a guided wavelength,
$\lambda_g/2 = c/(2f\sqrt{\varepsilon_\mathrm{eff}})$, where $c$ is the speed of light
and $\varepsilon_\mathrm{eff}$ the effective permittivity~\cite{pozar2011,balanis2016}.
For a 3~mm strip on this substrate, $W/h = 1.97$, and the quasi-static microstrip
expression gives $\varepsilon_\mathrm{eff} \approx 2.36$~\cite{pozar2011}. Half a guided
wavelength is then about 98~mm at 1~GHz and 16~mm at 6~GHz. A path of two or three
patch pitches therefore resonates near the upper edge of the band, while a resonance
near the lower edge needs a path as long as the full 99~mm width of the lattice. Fig.~\ref{fig:hardware} shows the top-down view and the substrate stack.
\vspace{2mm}

A switchable copper bridge of dimensions $1 \times 3.4$~mm is placed in each gap between two adjacent patches, modelled as ideal copper when on and as an open circuit when off. Treating the connections as ideal during design follows earlier pixel-antenna optimisation, where connections were represented as hardwires~\cite{shen2017}, and it isolates the effect of the switch pattern from that of any particular switch technology. The effect of practical switches is discussed in Section~\ref{sec:discussion}. Each patch is connected to the patch below by a vertical bridge, giving 15 bridges per column across 16 columns, for 240 vertical bridges. The full switch state is therefore a 480-element binary vector, and the design space has $2^{480} \approx 3.1 \times 10^{144}$ elements. For numerical processing, the switch state is stored as a tensor of shape $(2, 16, 16)$, where channel~0 holds the horizontal bridges and channel~1 holds the vertical bridges. The horizontal bridges naturally form a $15 \times 16$ array and the vertical bridges a $16 \times 15$ array, so the two sets do not share the same shape. \par
Each is padded to a full $16 \times 16$ grid so that both can be stored together as two channels and each bridge stays aligned with the patch it belongs to. The cost of this alignment is one unused column in channel~0 and one unused row in channel~1, which are zero for every pattern. A single port, P1, is connected to the patch at position $(1, 8)$ in 1-based indexing, which is the leftmost column, eighth row. The port consists of a tapered microstrip feed of 16~mm length, connected to a waveguide port at its outer edge, with the taper providing an impedance match between the port boundary and the patch interface near 3.5~GHz. The active bridges form a single connected metallic region that contains P1, and this region radiates and produces one or more resonant dips in $S_{\mathrm{11}}$. Every pattern used in this work is a single connected component rooted at P1, so the reflection coefficient at P1 is the only observable utilized by the design method.

%=====================================================================
\section{Dataset Generation}
\label{sec:dataset}

The forward model is trained on switch patterns paired with their simulated
reflection coefficients, so a dataset has to be built by simulation before any model
can be fitted. Which patterns enter that dataset is itself a design decision, since
the $2^{480}$ configuration space cannot be sampled exhaustively and every candidate
costs a full-wave simulation. Training patterns are therefore produced by a
stochastic generator in which one of eight pattern families is selected with equal probability.
The families are a square region grown around P1, a rectangular region, a ring, a
split-ring resonator, a small patch cluster with thin branches, a fractal, a
breadth-first random blob from P1, and a hybrid combining a square base with
breadth-first growth. Every family is anchored at P1. Each candidate pattern is
validated against the connectivity constraint, which requires the active bridges to
form a single connected region containing P1. Patterns that fail validation are
rejected and regenerated.
\vspace{2mm}
Shape families are used rather than uniform random switch states because uniform
sampling does not produce usable patterns. To confirm this, 20{,}000 switch vectors
were drawn at each of several values of an activation probability $p$, with each of
the 480 bits set independently. A draw was counted as valid if every active bridge
lay inside a single connected region containing P1, which is the condition satisfied
by every pattern in the dataset. At $p = 0.12$, which activates 57.6 bridges on
average and therefore matches the dataset mean of 58.6 closely, not one valid pattern
appeared in 20{,}000 draws: the region reachable from P1 contained on average 1.5
patches, while almost all of the remaining switches lay in disconnected islands isolated from the feed port. The fragmentation persists as $p$ increases. Between $p = 0.02$
and $p = 0.30$ the mean number of active bridges rises from 9.6 to 144, yet the mean
region reachable from P1 grows only from 1.1 to 4.1 patches, so adding metal does not
add radiating structure. Connectivity is reached only once the lattice is close to
saturated: at $p = 0.6$ the mean region covers 180 of the 256 patches, which is an
almost solid sheet rather than a shaped resonator, and its size can no longer be
chosen. The pattern size is also fixed by uniform sampling, since the number of active
bridges is binomial with standard deviation $\sqrt{480\,p(1-p)}$, which is about 7
switches at $p = 0.12$, whereas the dataset must span three to several hundred
switches to place resonances across the full band. The connected region therefore has
to be constructed rather than sampled.

% Can use this, from uniform spacing to sampled: paragraph above,... just different way of explaing the same thing,.. Independent sampling also removes control over the pattern size. For a fixed $p$
% the number of active bridges is binomial with standard deviation
% $\sqrt{480\,p(1-p)}$, which is about 7 switches at $p = 0.12$, so nearly every
% draw contains between 43 and 72 active bridges. Since the resonant frequency of a
% connected region is set by its current path length, a dataset confined to one
% narrow band of sizes would cover only a narrow band of frequencies, whereas the
% patterns used here span 3 to 409 active switches.

\begin{figure}[!t]
\centering
\includegraphics[width=\columnwidth]{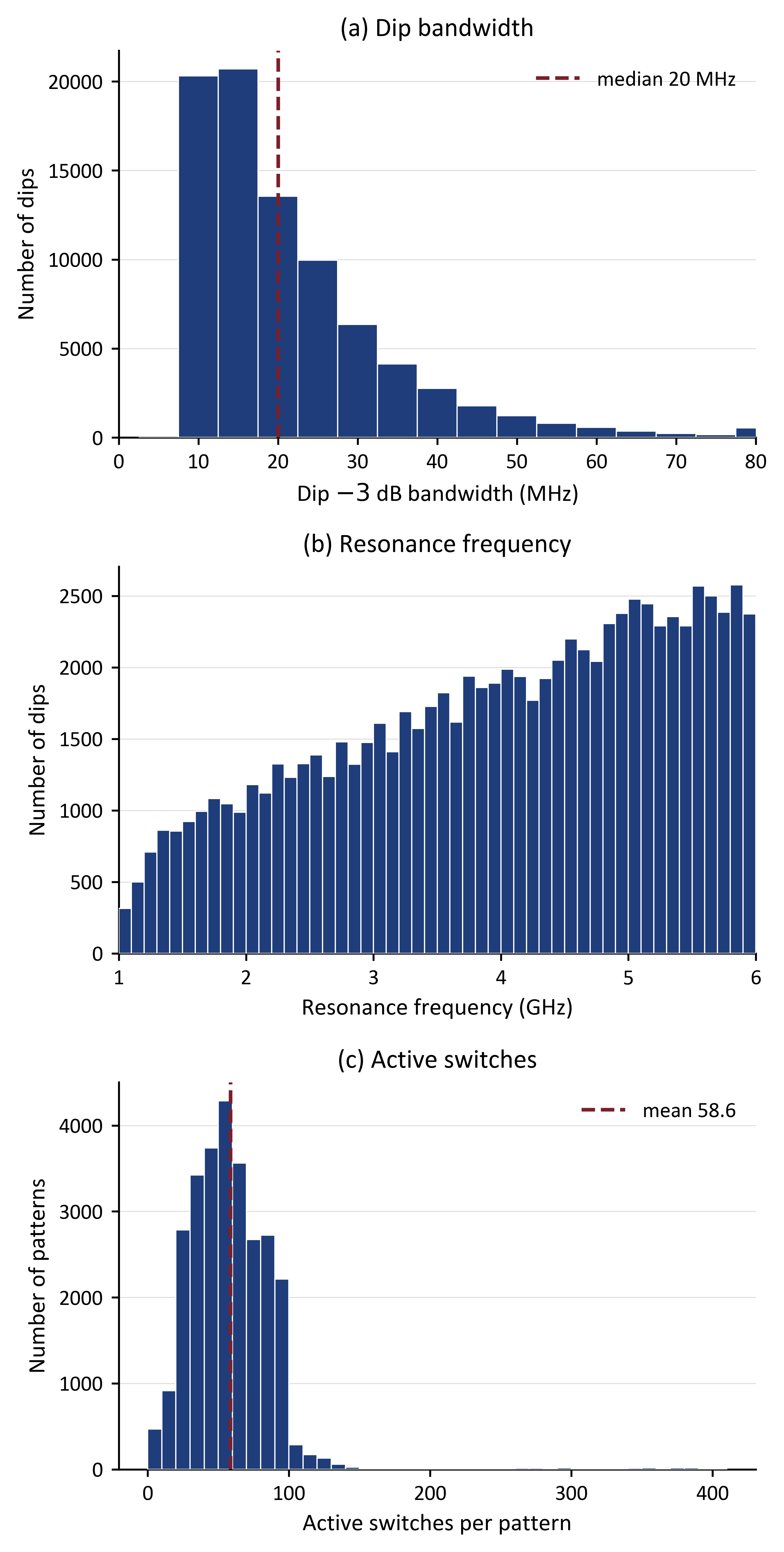}
\caption{Dataset statistics. (a) Distribution of the $-3$~dB bandwidth of detected
dips, with a median of 20~MHz. (b) Distribution of resonance frequency across all
detected dips. (c) Distribution of the number of active switches per pattern.}
\label{fig:datasetstats}
\end{figure}

A pattern is described by two related quantities, both of which are used later in
this work. The first is the number of active switches, meaning the horizontal and
vertical bridges turned on between adjacent patches. This is the quantity the design
variable controls directly, since the 480-bit vector is a list of switch states. The
second is the number of patches joined into the P1-connected region, which is the
metal that carries current and radiates. Turning on a switch either attaches a
further patch to the region or closes a loop inside it, so the two counts rise
together without being equal. Across the dataset the mean is 58.6 active switches with a range from 3 to 409, and
the connected region ranges from 4 patches to 225, with half the dataset lying
between 26 and 48 patches. The distribution of active switch counts is shown in
Fig.~\ref{fig:datasetstats}(c). The number of
independent loops in a region, given by the number of switches minus the number of
patches plus one, averages 21.7 and is used as one of the graph features described in
Section~\ref{sec:forward}.

Each generated pattern is simulated using the time-domain transient solver with open
boundary conditions in CST Microwave Studio~\cite{cst2023}. Copper and substrate are
modelled as lossy materials with the parameters given in Section~\ref{sec:hardware}.
The frequency sweep covers 1.0 to 6.0~GHz with 1001 uniformly spaced samples, giving
a resolution of 5~MHz. The reflection coefficient $S_{\mathrm{11}}$ at P1 is the observable metric, used throughout this
work.

A total of 31{,}659 patterns were generated and simulated. Of these, 27{,}730
patterns, or 87.6\%, produced an $S_{\mathrm{11}}$ curve with a minimum value below the
$-10$~dB threshold, which is the conventional return loss level at which an antenna
is considered matched. In a further step, only patterns with at least one local
minimum below $-10$~dB that also has a prominence of at least 1~dB were retained, which
distinguishes a true resonance from a curve that only marginally crosses the threshold. In this step 39 patterns were removed whose $S_{\mathrm{11}}$ touched $-10$~dB at isolated noisy points
without forming a proper resonance, leaving a working dataset of 27{,}691 patterns
containing 83{,}482 detected resonances in total.

The resonances in this dataset are narrowband. As shown in
Fig.~\ref{fig:datasetstats}(a), the median $-3$~dB bandwidth of an individual dip is
20~MHz, which is a small fraction of the 5000~MHz sweep. The narrowest resonances sit
at 10~MHz, which is the resolution limit of the frequency sweep, and the widest
reaches 230~MHz. Each pattern has on average 3.01 deep dips. The deepest dip of a
pattern has a mean depth of $-21.8$~dB, while the mean taken across all detected dips
is $-17.3$~dB. Due to this narrowbandness property the classification
approach is utilised, since a resonance that occupies so
little of the frequency axis is easily missed by a model trained to reproduce the
full curve, described in Section~\ref{sec:forward}. The distribution of resonance frequencies is shown in
Fig.~\ref{fig:datasetstats}(b). The 1.0 to 1.5~GHz interval contains 3{,}240 dips and
the 5.5 to 6.0~GHz interval contains 12{,}406, so resonances are present across the
full band with higher density at the upper end, which is consistent with the
geometric tendency of small connected regions to resonate at higher frequencies. The dataset is partitioned into training, validation, and test subsets in the ratio
$80{:}10{:}10$ using a pseudo-random permutation with a fixed seed of 42. 

This gives
22{,}152 training patterns, 2{,}769 validation patterns, and 2{,}770 test patterns.
The same partition is used for all subsequent experiments. The summary of the dataset is shown in Table~\ref{tab:dataset}.

\begin{table}[!t]
\caption{Dataset summary.}
\label{tab:dataset}
\centering
\begin{tabular}{lr}
\toprule
\textbf{Quantity} & \textbf{Value} \\
\midrule
Patterns simulated in CST               & 31{,}659 \\
Patterns after $S_{\mathrm{11}} < -10$~dB filter & 27{,}730 \\
Patterns after prominence filter        & 27{,}691 \\
\hspace{1em}Training subset             & 22{,}152 \\
\hspace{1em}Validation subset           &  2{,}769 \\
\hspace{1em}Test subset                 &  2{,}770 \\
Mean active switches per pattern        & 58.6 \\
Range of active switches                & 3 to 409 \\
Mean patches connected to P1            & 38 \\
Range of connected patches              & 4 to 225 \\
Mean deep dips per pattern              & 3.01 \\
Total deep dips detected                & 83{,}482 \\
Mean depth of deepest dip per pattern   & $-21.8$~dB \\
Mean depth across all dips              & $-17.3$~dB \\
Median dip $-3$~dB bandwidth            & 20~MHz \\
Range of dip $-3$~dB bandwidth          & 10 to 230~MHz \\
Frequency-sweep range                   & 1.0 to 6.0~GHz \\
Frequency-sweep samples                 & 1001 \\
\bottomrule
\end{tabular}
\end{table}

% =====================================================================
\vspace{-1.5mm}
\section{Forward Surrogate Model}

\label{sec:forward}

Regression of the full reflection-coefficient curve was examined first, on patterns
produced by the same generator and simulation protocol. A convolutional model trained
to reproduce $|S_{11}|$ at all 1001 frequency points reached a mean absolute error of
1.2~dB, which on its own suggests an accurate model. Its predicted resonance, however,
lay on average 1.46~GHz from the simulated one, nearly a third of the 5~GHz band.
The low error was earned on the flat baseline, which covers most of the sweep, while
the resonance, the feature most relevant to inverse design, was not located.

This motivates the central change made in this work. The forward problem is
reformulated from regression of the curve to multi-label classification of resonance
location. Instead of reproducing $|S_{11}|$ at every frequency point, the model
predicts, for each of 100 frequency bins, the probability that a deep resonance is
present. The training signal is thereby concentrated on the resonances rather than spread over the baseline. This allows the model to focus on the feature that matters for inverse design and provide the inverse search with a direct estimate of where a given pattern resonates.

\subsection{Soft Label Construction}
\vspace{2mm}
The forward model is trained to predict, from a switch pattern, whether a deep
resonance is present in each of 100 frequency bins. The 1.0 to 6.0~GHz band is uniformly partitioned
into 100 bins, so each bin is 50~MHz wide. The label vector for a pattern
is built from its CST-simulated $S_{\mathrm{11}}$ curve by detecting local minima.

A local minimum at frequency $f_d$ is treated as a deep resonance if its depth is
below $-10$~dB and its prominence is at least 1~dB. The prominence is the vertical distance from the minimum up to the lower of the two adjacent local maxima, so a shallow dip lying on the shoulder of a deeper one is correctly assigned a small prominence and rejected. The $-3$~dB bandwidth $\mathrm{BW}_d$ is the width of the
frequency interval over which the curve remains within 3~dB of the value at the
minimum. A Gaussian bump is added to the label vector for each detected resonance,
with standard deviation
\begin{equation}
\sigma_d = \max\!\left(\frac{\mathrm{BW}_d}{2}, \; \sigma_{\min}\right),
\label{eq:sigma}
\end{equation}
where $\sigma_{\min} = 40$~MHz.
The floor $\sigma_{\min}$ is the operative term for almost every resonance in this
dataset. The bandwidth term exceeds it only when $\mathrm{BW}_d > 80$~MHz, which
holds for 0.5\% of detected dips, so 99.5\% of labels are Gaussians of
fixed 40~MHz width. This is deliberate. With a median dip bandwidth of 20~MHz and a
bin width of 50~MHz, a label scaled directly to the resonance bandwidth would be
narrower than a single bin and would place nearly all of its mass on that one bin,
leaving no gradient signal in the neighbouring bins. \par
\begin{figure}[!t]
\centering
\includegraphics[width=\columnwidth]{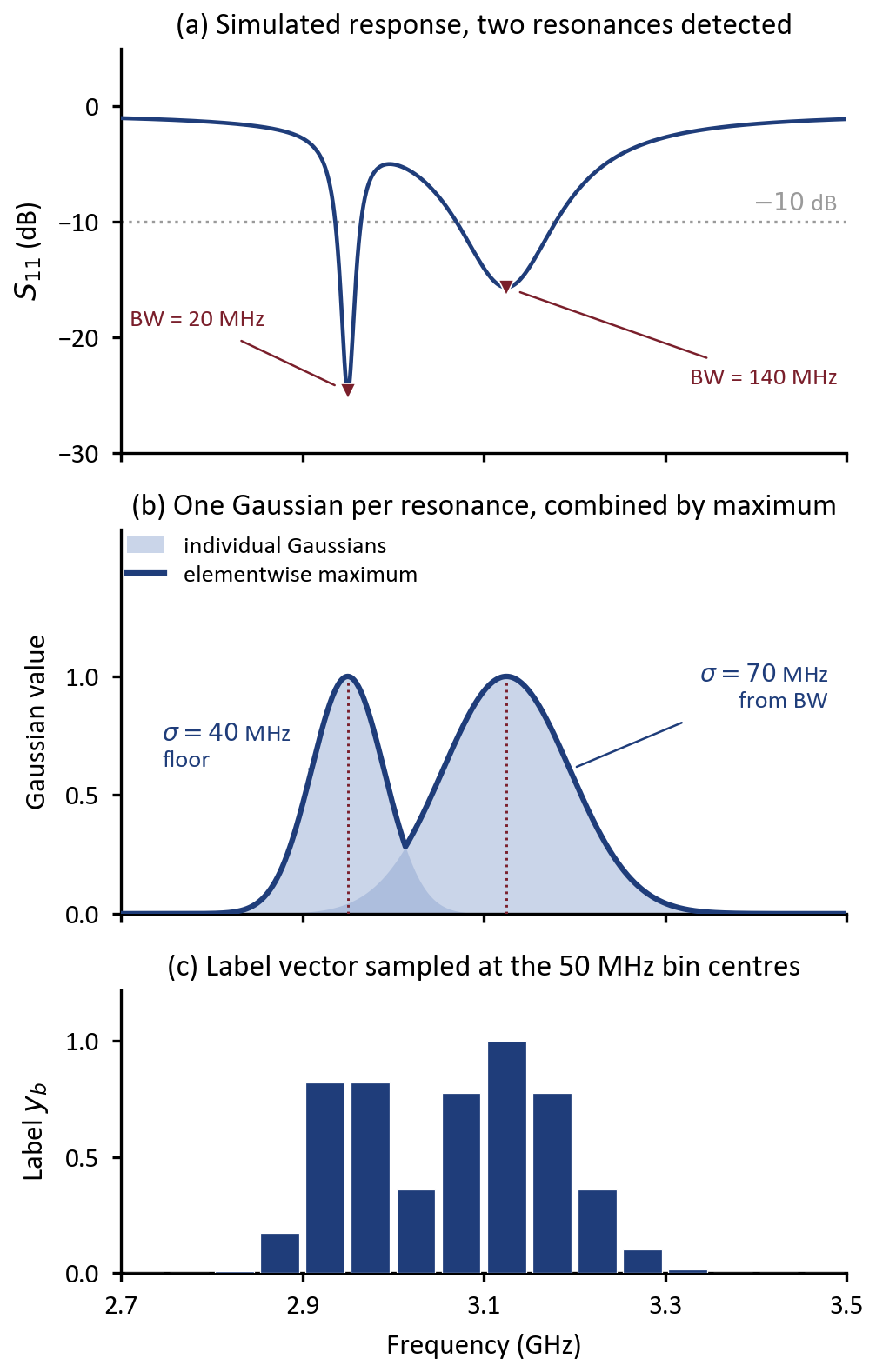}
\caption{Soft-label construction, shown on an illustrative response. (a) Two
resonances are detected, one of 20~MHz and one of 140~MHz bandwidth. (b) Each
resonance contributes a Gaussian whose width follows~(\ref{eq:sigma}); the narrow
resonance takes the 40~MHz floor while the broad one takes half its bandwidth, and
where the two overlap the elementwise maximum of~(\ref{eq:soft}) is retained.
(c) The combined field sampled at the 100 bin centres gives the label vector used
for training. The curve in (a) is drawn for illustration; the bin width, the
$\sigma$ floor, and the combination rule are those used throughout.}
\label{fig:softlabels}
\end{figure}

 Through this floor, every resonance, however narrow, is guaranteed to be spread across several adjacent bins, while the
bandwidth term widens the labels of the small minority of broad resonances. When a pattern has more than one detected resonance, the Gaussian contributions are
combined by elementwise maximum, so the nearest resonance determines the label at
each bin. The label at bin $b$, whose centre frequency is $f_b$, is
\begin{equation}
y_b = \max_d \exp\!\left(-\frac{(f_b - f_d)^2}{2\sigma_d^2}\right),
\label{eq:soft}
\end{equation}
where the maximum is taken over all detected resonances $d$ in the pattern and
$\sigma_d$ is given by~(\ref{eq:sigma}). The bin nearest a true resonance therefore
has $y_b$ close to unity, neighbouring bins receive partial credit, and bins far from
any detected resonance have $y_b$ close to zero. A label attains unity exactly when a
resonance coincides with a bin centre, which occurs for about one dip in ten because
the bin centres and the frequency sweep share the same 5~MHz grid. The construction
is illustrated in Fig.~\ref{fig:softlabels}.

\subsection{Input Representation}
The forward model has two inputs. The first is an image tensor of shape
$(3, 16, 16)$. The horizontal-bridge state is held in channel~0, the
vertical-bridge state in channel~1, and a binary mask marking the patches
connected to P1 in channel~2. The port-connectivity information is thereby made
directly available to the convolutional layers, rather than having to be derived from the bridge channels alone.

The second input is a vector of five scalar features describing the P1-connected
subgraph: the number of patches in the component, the number of bridges inside the
component, the longest hop distance from P1 measured in graph edges, the number of
independent cycles in the component computed as edges minus nodes plus one, and the
area of the bounding box enclosing the connected patches. The connectivity mask and
the five features are obtained from a single breadth-first traversal of the bridge
graph starting at P1, in time linear in the number of active bridges. The features
are standardized using the mean and standard deviation of the training subset.

\subsection{Network Architecture}
\begin{figure}[!t]
\centering
\includegraphics[width=\columnwidth]{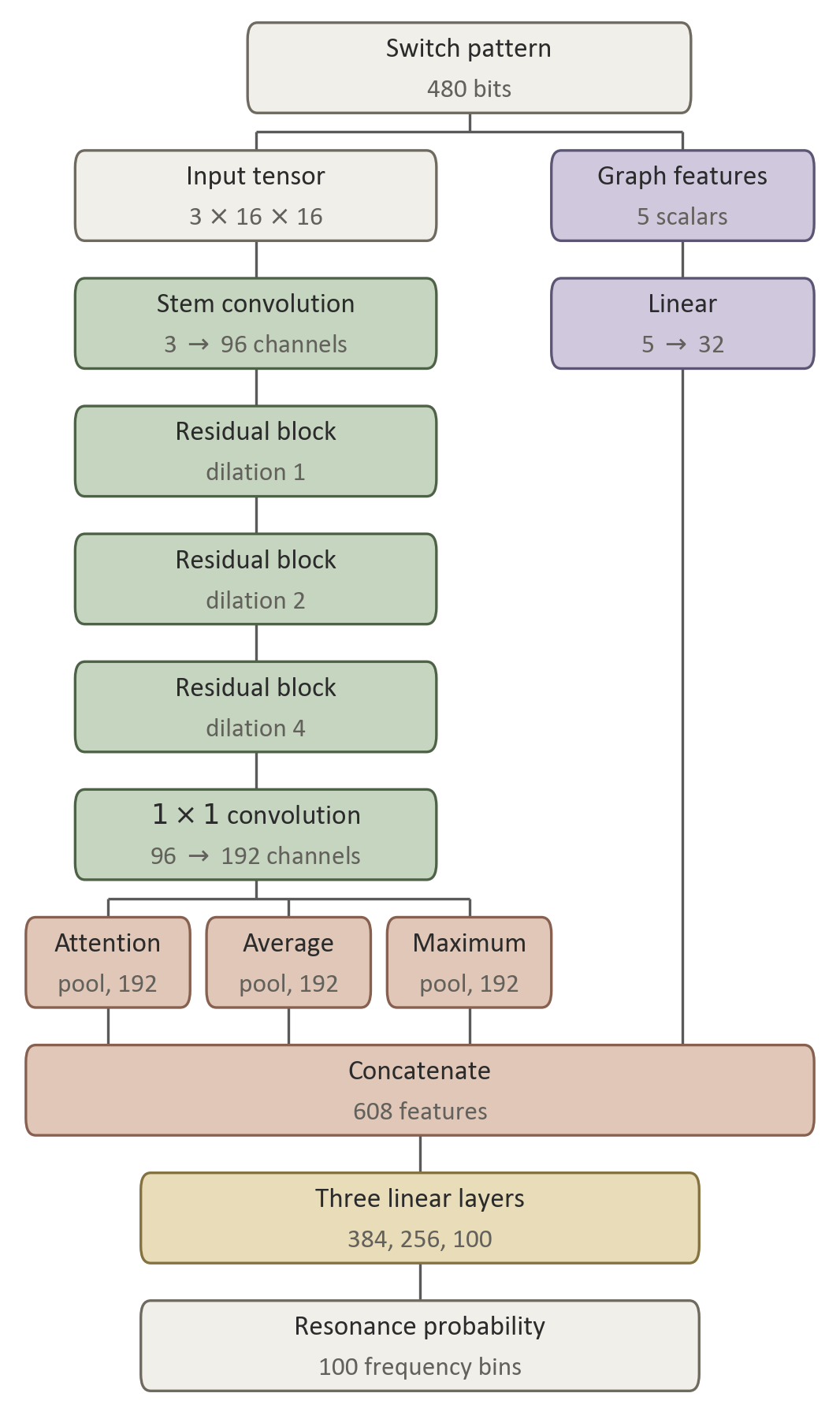}
\caption{Forward surrogate network architecture. The image branch processes the
$(3, 16, 16)$ input through a stem and three dilated residual blocks. Three pooling
operations and a graph-feature layer produce four descriptors that are concatenated
and passed through a three-layer head to yield 100 frequency-bin logits.}
\label{fig:architecture}
\end{figure}

The forward model is a convolutional network with an auxiliary scalar branch. The architecture is shown in
Fig.~\ref{fig:architecture} and its hyperparameters are listed in Table~\ref{tab:hyperparams}.

The image branch begins with a stem convolution, in which the 3 input channels are
expanded to 96 feature channels. The stem is followed by three residual blocks~\cite{he2016} with
dilation rates 1, 2, and 4, each block containing two dilated convolutions with batch
normalization~\cite{ioffe2015} and the GELU activation~\cite{hendrycks2016}. The receptive field is expanded by dilated convolutions~\cite{yu2016} without
downsampling the feature map, so the full spatial resolution is preserved throughout. With a kernel
size of 3, the cumulative receptive field is $3 \times 3$ after the stem,
$7 \times 7$ after the first block, $15 \times 15$ after the second, and
$31 \times 31$ after the third. \par The second block therefore already spans almost the
entire lattice and the third covers it with margin, thereby providing a hierarchical representation of local, mid-range, and full-board structures whose
spatial extent is only 16 cells. The feature map is then projected to 192 channels by
a $1 \times 1$ convolution. The 192-channel feature map is summarized using three parallel pooling operations. In learnable attention pooling~\cite{vaswani2017}, a soft mask is computed over the $16 \times 16$ spatial positions, followed by a weighted sum of the feature vectors. Global average pooling computes the mean across spatial positions, while global maximum pooling extracts the maximum activation for each channel. These operations provide complementary representations: attention pooling produces a weighted spatial representation, average pooling captures the mean activation level, and maximum pooling retains the strongest response in each channel. \par Each pooling operation produces a 192-dimensional descriptor. The five-feature scalar input is passed through a single linear layer with GELU
activation that maps it to 32 features. The four descriptors, three pools of 192 each
together with the 32 graph features, are concatenated into a 608-dimensional vector
and passed through a head of three linear layers with hidden sizes 384 and 256 and an
output of 100 logits. Dropout is applied after each of the two hidden layers. The
network has 879{,}397 trainable parameters.

\vspace{-3.5mm}
\subsection{Training Procedure}
The output of the model is a vector of 100 logits, one per frequency bin. A sigmoid
is applied to each logit to obtain the predicted probability that a deep resonance is
present in that bin. The model is trained with a per-bin weighted binary cross-entropy
loss. \par Let $n$ index the training sample within a minibatch of size $N$, let $b$ index
the frequency bin out of $B = 100$, let $z_{n,b}$ be the model logit, let $y_{n,b}$ be
the soft label defined by~(\ref{eq:soft}), let $\sigma(\cdot)$ denote the logistic
sigmoid, and let $w_b$ be a per-bin weight. The loss is
\begin{equation}
\begin{split}
\mathcal{L} = -\frac{1}{NB} \sum_{n} \sum_{b} \Big[ &w_b \, y_{n,b} \log \sigma(z_{n,b}) \\
+\; &(1 - y_{n,b}) \log(1 - \sigma(z_{n,b})) \Big].
\end{split}
\label{eq:bce}
\end{equation}
The first term inside the sum is the cost incurred for assigning low probability to a
bin that contains a resonance. The second term is the cost incurred for assigning high
probability to a bin that does not. The positive-class term is scaled by the factor $w_b$ to
compensate for class imbalance. Low-frequency bins have far fewer positive examples
than high-frequency bins, so without compensation the loss would be dominated by the
more abundant high-frequency positives.

The per-bin weight is computed from the marginal positive-class frequency of the
training subset as
\begin{equation}
w_b = \frac{1 - p_b}{p_b}, \qquad p_b = \frac{1}{|\mathcal{D}_\text{train}|} \sum_{n \in \mathcal{D}_\text{train}} \mathbb{I}[y_{n,b} > 0.5],
\label{eq:posweight}
\end{equation}
where $p_b$ is the fraction of training patterns whose label at bin $b$ exceeds 0.5
and $\mathbb{I}[\cdot]$ is the indicator function. The form of $w_b$
in~(\ref{eq:posweight}) is the standard positive-class weighting used with binary
cross-entropy under class imbalance: the expected gradient contribution from positive and negative examples is
equalised in each bin. A large $w_b$ is assigned to bins with $p_b$ close to zero,
while bins with $p_b$ close to 0.5 are assigned $w_b$ close to unity. Before
the weight is formed, $p_b$ is clamped to the interval $[10^{-4}, 1 - 10^{-4}]$, which
bounds $w_b$ at $10^4$ and avoids a division by zero in any bin that contains no
positive example in the training subset.

The network is trained with the AdamW optimizer~\cite{loshchilov2019}, a variant of
Adam~\cite{kingma2015} with decoupled weight decay, at an initial learning rate of
$10^{-3}$, weight decay of $10^{-5}$, and minibatch size 128. The learning rate is
annealed by a cosine schedule~\cite{loshchilov2017} to a floor of $10^{-5}$ over a
maximum of 150 epochs. Mixed-precision training is used. Early stopping is applied on
the validation mean area under the receiver operating characteristic curve with a
patience of 15 epochs. The best validation checkpoint, reached at epoch 53, is
retained for all subsequent evaluations.

\begin{table}[!t]
\caption{Forward model hyperparameters.}
\label{tab:hyperparams}
\centering
\begin{tabular}{ll}
\toprule
\textbf{Component} & \textbf{Value} \\
\midrule
Image input shape           & $(3, 16, 16)$ \\
Graph-feature input         & 5 scalars \\
Stem output channels        & 96 \\
Residual blocks             & 3 (dilations 1, 2, 4) \\
Convolutions per block      & 2 \\
$1 \times 1$ projection     & 96 to 192 channels \\
Pooling                     & Attention + GAP + GMP \\
Graph-feature layer         & 5 to 32 \\
Concatenated feature dim    & 608 \\
Head hidden sizes           & 384, 256 \\
Output dim                  & 100 \\
Dropout (head)              & 0.25 \\
Activation                  & GELU \\
Normalization               & BatchNorm2d \\
Trainable parameters        & 879{,}397 \\
\midrule
Optimizer                   & AdamW \\
Initial learning rate       & $10^{-3}$ \\
Learning-rate schedule      & Cosine to $10^{-5}$ \\
Weight decay                & $10^{-5}$ \\
Minibatch size              & 128 \\
Maximum epochs              & 150 \\
Early-stopping patience     & 15 epochs \\
Loss                        & Weighted BCE, (\ref{eq:bce}) \\
\bottomrule
\end{tabular}
\end{table}

\section{Inverse Design Methodology}
\label{sec:inverse}

\subsection{Search Space and Validity Constraint}
The inverse-design task is a constrained discrete optimization over the 480-bit
switch space. A pattern is valid when its active bridges form a single connected
region that contains P1 and holds at least two patches. The reason for the
constraint is that the response is measured at P1, so any metal not joined to P1
carries no fed current and does not contribute to the reflection coefficient. Every
pattern in the training set has this property, as reported in
Section~\ref{sec:dataset}, so restricting the search in the same way also keeps the
candidates inside the distribution the forward model was trained on.

The constraint is maintained throughout the search rather than applied as a filter
afterwards. Moves that add a bridge preserve connectivity by construction, and moves
that remove a bridge are checked before they are accepted, so every state the chain
visits is a valid pattern.

\subsection{Objective Function}
Let $s \in \{0, 1\}^{480}$ denote a switch state, let $f_t$ be the requested target
frequency, and let $b_t$ be the index of the 50~MHz bin that contains $f_t$. Let
$P(s; b) \in [0, 1]$ be the probability that bin $b$ contains a deep resonance, as
predicted by the trained forward model of Section~\ref{sec:forward} after applying
the sigmoid to the corresponding logit. Let $n(s)$ be the number of patches in the
P1-connected region of $s$. The objective maximized by the inverse search is
\begin{equation}
J(s) = \max_{b \in \{b_t - 1, \, b_t, \, b_t + 1\}} P(s; b) \;-\; \lambda \cdot \frac{n(s)}{256}.
\label{eq:objective}
\end{equation}

The first term is a localised score. The highest predicted resonance probability within a three-bin window of
$\pm 50$~MHz around the target is returned. The window
absorbs small frequency offsets between the search and the target, which would
otherwise penalize a candidate whose resonance falls just outside the centre bin.

Large connected regions are penalized by the second term. The denominator 256 is the total
number of patches on the board, so $n(s)/256$ lies between 0 and 1. The penalty
coefficient is $\lambda = 0.25$, which reduces the score by about $10^{-3}$ for each
additional patch. That is small compared with the score range of $[0, 1]$ but large
enough to separate candidates of otherwise similar score.

A hard cap of $n_{\max} = 80$ patches is also applied while moves are generated. The
cap is set by training coverage rather than by fabrication. The training subset
averages 38 connected patches per pattern, and only 0.7\% of the dataset
exceeds 150 patches, so a search allowed to drift well above that range would rely on extrapolated predictions from the forward model. The cap and the penalty together keep the
search inside the region of input space where the model has support.

\subsection{Simulated Annealing}
The search uses a simulated annealing chain~\cite{kirkpatrick1983} that operates
directly on the switch state. Each move changes one bridge. Three kinds of move are
used.
\begin{itemize}
\item A \textit{grow} move activates a bridge between a patch already in the region
and an adjacent patch that is not, which adds one patch to the region.
\item A \textit{close-loop} move activates a bridge between two patches that are both
already in the region, which adds a cycle but no patch.
\item A \textit{remove} move deactivates one active bridge chosen at random. The
resulting region is then tested, and the move is kept only if the region is still
connected to P1 and still holds at least two patches. Up to six bridges are tried in
turn, and if none of them can be removed the step is skipped.
\end{itemize}
At each step the kinds of move that have at least one candidate bridge available are
collected, and one kind is drawn uniformly from that set. Grow moves are excluded
once the region has reached $n_{\max}$ patches, so the region never exceeds the cap.

The proposed state is accepted or rejected by the Metropolis rule. Let
$J_\text{cur}$ be the objective value of the current state, $J_\text{new}$ that of
the proposed state, and $T$ the current temperature. The acceptance probability is
\begin{equation}
p_\text{acc} = \min\!\left(1, \, \exp\!\left(\frac{J_\text{new} - J_\text{cur}}{T}\right)\right).
\label{eq:metropolis}
\end{equation}
Improving moves are always accepted. Worsening moves are accepted with a probability
that falls as the gap $J_\text{new} - J_\text{cur}$ grows and as $T$ drops. Early in the chain, the relatively high $T$ increases the probability of accepting worsening moves, allowing transitions away from local maxima. As the $T$ decreases, the acceptance probability for worsening moves declines, and the search increasingly favours objective-improving transitions. The temperature follows a geometric schedule from $T_0 = 1.0$ to
$T_f = 0.01$ across the $S$ steps of the chain, and is reduced at every step whether
or not a move was made. For each target frequency, $M = 10$ chains are run independently from different
random seeds, each for $S = 5000$ steps. Each chain is initialised with a random
P1-rooted region of 4 to 20 patches. The best state seen during the run is returned by each chain rather than
its final state, so a poor move late in the chain cannot spoil the result. The seed of chain $m$ at target $f_t$ is $44 + 31m + \lceil 10 f_t
\rceil$, which gives a distinct seed for every chain and target.
Algorithm~\ref{alg:sa} summarises the procedure.
\vspace{-3.5mm}

\begin{algorithm}[!t]
\caption{Simulated annealing for inverse design.}
\label{alg:sa}
\begin{algorithmic}[1]
\REQUIRE target frequency $f_t$, chains $M$, steps $S$, temperatures $T_0$ and $T_f$, patch cap $n_\text{max}$, forward model probabilities $P(\cdot;\cdot)$.
\ENSURE $M$ candidate states.
\FOR{$m = 1$ \TO $M$}
  \STATE Initialise $s$ as a random P1-rooted region of 4 to 20 patches.
  \STATE $s_\text{best} \leftarrow s$
  \FOR{$t = 1$ \TO $S$}
    \STATE $T \leftarrow T_0 \cdot (T_f / T_0)^{t / S}$
    \STATE Collect the available move kinds at $s$, excluding grow if $n(s) = n_\text{max}$.
    \STATE Draw one kind uniformly and apply a move of that kind to obtain $s'$.
    \IF{no valid move was found}
      \STATE \textbf{continue}
    \ENDIF
    \STATE $\Delta \leftarrow J(s') - J(s)$ from (\ref{eq:objective})
    \IF{$\Delta > 0$ \OR $\mathrm{rand}() < \exp(\Delta / T)$}
      \STATE $s \leftarrow s'$
      \IF{$J(s) > J(s_\text{best})$}
        \STATE $s_\text{best} \leftarrow s$
      \ENDIF
    \ENDIF
  \ENDFOR
  \STATE $c_m \leftarrow s_\text{best}$
\ENDFOR
\STATE Order $\{c_1, \dots, c_M\}$ by greedy max-min Hamming distance.
\RETURN the ordered candidates.
\end{algorithmic}
\end{algorithm}

\subsection{Candidate Ordering}
% \vspace{-1.5 mm}
The candidate states returned for a given target frequency are often variations on a single topology.
They are therefore ordered by greedy max-min selection before being passed to
full-wave simulation. The first candidate is the one with the highest objective
value. Each following candidate is the one whose smallest Hamming distance to the
candidates already chosen is largest. The most distinct topologies are placed first by the ordering, so that a run that simulates fewer candidates than the chains
produced still covers several regions of the pattern space. In the results reported
here the number of candidates simulated per target equals the number of chains, so
all ten outputs were retained.
% =====================================================================
% =====================================================================
% =====================================================================
\section{Results}
\label{sec:results}
% =====================================================================
\vspace{2mm}
\subsection{Forward Model Performance}
% \vspace{-2mm}
The forward model is evaluated on the held-out test set. Two metrics are reported. The first is the area under the receiver operating characteristic curve (AUC),
computed for each frequency bin. For a fixed bin $b$, AUC is the probability that a
randomly chosen pattern with a real resonance in bin $b$ receives a higher predicted
score than a randomly chosen pattern without one. A binary target is obtained by
thresholding the soft label at 0.5. AUC does not depend on a decision threshold. The
mean per-bin AUC is 0.737, against a chance value of 0.5, and no bin falls below
chance.

The second metric is precision at $K$ (P@K). For a fixed target bin, P@K is the
fraction of the $K$ patterns with the highest predicted score at that bin that
actually contain a resonance in the bin. This matches the way the model is used
inside the inverse search, where candidates are ranked by their predicted score at
the target bin and only the top-ranked ones are taken forward. P@K is averaged over
20 evenly spaced target bins with centre frequencies from 1.125 to 5.875~GHz. The
values of P@1, P@5, P@10, and P@20 are 0.550, 0.470, 0.405, and 0.393. Precision
falls as $K$ grows, as expected when a ranked list is extended.

\begin{figure}[t!]
\centering
\includegraphics[width=\columnwidth]{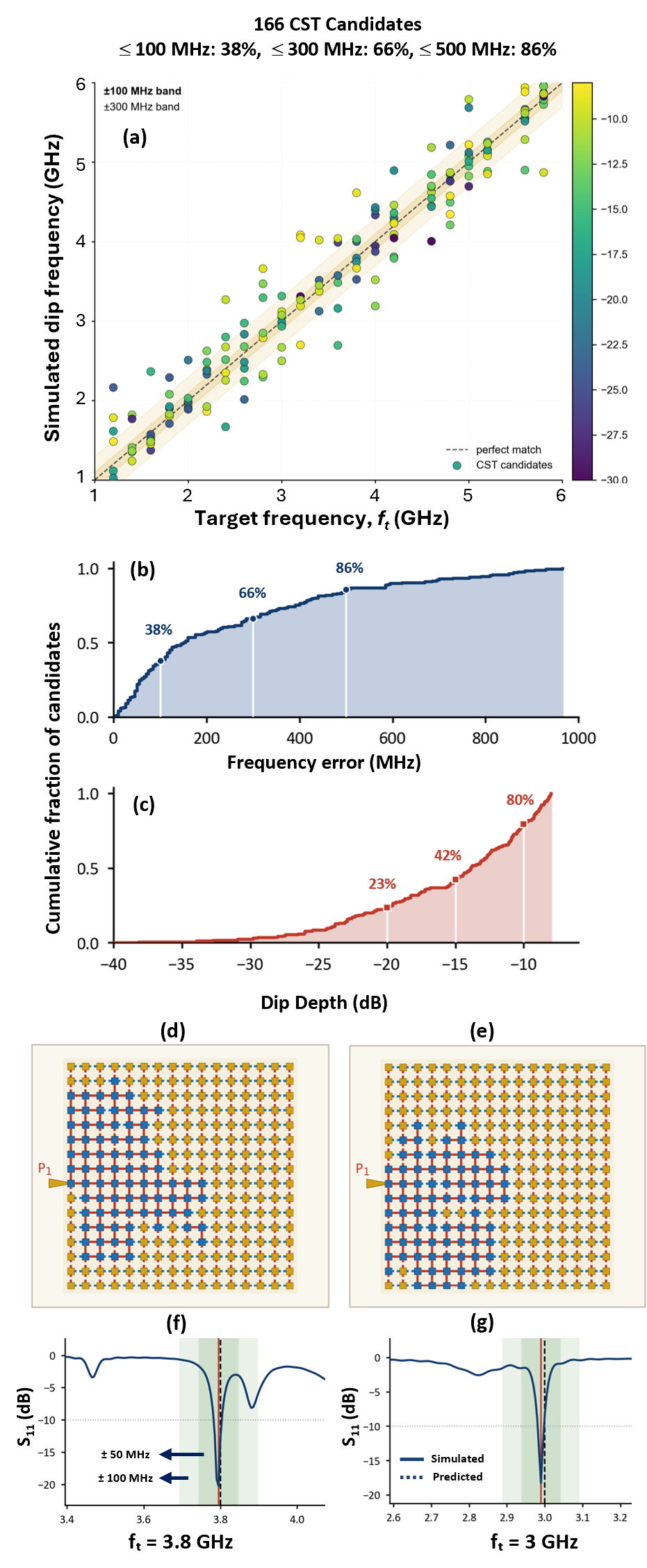}
\caption{Inverse design results. (a) Requested against simulated dip frequency for
the 166 successful candidates. (b) and (c) Cumulative distributions of the
frequency error and the dip depth. (d) and (e) Two generated switch patterns.
(f) and (g) Their simulated reflection coefficients.}
\label{fig:scatter}
\end{figure}

\subsection{Pipeline Verification by Full-Wave Simulation}
The complete pipeline is verified by full-wave simulation of 220 generated
candidates. Twenty-two target frequencies between 1.2 and 5.8~GHz are used, with 10
candidates generated per target by the inverse search of
Section~\ref{sec:inverse}. All 220 candidates are simulated in CST under the same
protocol as the training dataset described in Section~\ref{sec:dataset}. Two quantities are measured for each candidate: where its resonance falls relative
to the requested frequency, and how deep that resonance is. Measuring the first
requires the resonance to be located, and a detection threshold of $-8$~dB is used
for that purpose. This is deliberately looser than the $-10$~dB matching criterion
applied to the training dataset in Section~\ref{sec:dataset}. A candidate whose $S_{\mathrm{11}}$ dips
to $-9$~dB at exactly the requested frequency has placed its resonance correctly,
and discarding it would hide a success of the search rather than reveal a failure.
The $-10$~dB criterion is retained as the matching criterion and is reported separately as the depth statistic below, so that frequency placement and match quality are assessed independently. Furthermore, each of the frequency accuracy figures quoted below is changed by no more than
two percentage points when the analysis is restricted to the candidates that
reach $-10$~dB.

For each simulated curve, a search window of $\pm 1$~GHz is placed around the
candidate's target frequency, and the deepest local minimum below $-8$~dB inside
that window is recorded as the produced resonance. A candidate is counted as successful when such a resonance is found. Of the 220 candidates, 166 are successful, a rate of
75\%.

Frequency accuracy is reported over these 166 candidates.
Fig.~\ref{fig:scatter}(a) shows the requested target frequency against the simulated
dip frequency for each one, with the dashed line marking a perfect match, the shaded
bands marking the $\pm 100$~MHz and $\pm 300$~MHz tolerance regions, and the marker
colour giving the depth of the resonance. Of these candidates, 38\% produce a
resonance within 100~MHz of the requested target, 66\% within 300~MHz, and
86\% within 500~MHz, with a median error of 152~MHz.
Fig.~\ref{fig:scatter}(b) gives the cumulative distribution of the frequency error,
read as the fraction of candidates whose dip falls within a given distance of the
request. These figures treat every candidate as a separate result. \par 
The method is not used in
that way: ten candidates are returned for each request, all ten are simulated, and the
one whose simulated dip lies closest to the request is taken as the answer. On that
basis the median error across the 22 requests is 23~MHz, as reported in
Section~\ref{sec:results}-C. The choice is made on the simulated response rather than
on the surrogate prediction, so no candidate is set aside on the strength of a model
output.

The depths of the 166 resonances range from $-8.0$ to $-38.2$~dB, and their
cumulative distribution is given in Fig.~\ref{fig:scatter}(c). Of these candidates,
80\% reach a depth of $-10$~dB or deeper and 42\% reach $-15$~dB or
deeper, with a median of $-14$~dB. Using the $-10$ dB matching criterion alone, 132 of the 220 simulated candidates, or 60\%, produce a matched
resonance within the search window. The colour scale in Fig.~\ref{fig:scatter}(a) shows that these deeper resonances
occur at target frequencies across the whole band rather than being confined to one
part of it.

Two generated designs are shown in Fig.~\ref{fig:scatter}(d) and (e), for requested
frequencies of 3.8 and 3.0~GHz. Active patches are drawn in blue, closed switches as
solid segments, and open switches with a break at the centre.
Fig.~\ref{fig:scatter}(f) and (g) give the corresponding simulated reflection
coefficients. In each, the shaded bands mark $\pm 50$~MHz and $\pm 100$~MHz about
the simulated dip. The solid vertical line marks the center dip of simulated $S_{\mathrm{11}}$, and the dashed vertical
line marks the resonance frequency predicted by the surrogate.

\subsection{Comparison with a Solver-Guided Search}
The solver is replaced by the surrogate inside the search, so the key issue is the resulting performance loss relative to a search that used the solver directly. To answer it, the
annealing chain of Section~\ref{sec:inverse} was rerun with the scoring function
of~(\ref{eq:objective}) replaced by a full-wave simulation at every step. The move
set, the connectivity constraint, the patch cap, and the temperature schedule are
unchanged; only the quantity being maximised differs. Each candidate is simulated
in CST and scored by the depth of the deepest point within $\pm 50$~MHz of the
target, so both searches reward a resonance near the target, one through its
simulated depth and the other through its predicted probability.

A matched comparison at the search length used here is not possible. In the surrogate-guided search 50{,}000 states are evaluated per request, and
simulating each one would take several months of solver time for a single target. The solver-guided
search was therefore given 200 simulations per target, which at an average of about 3.5~minutes per simulation, including data export, is about 11.5~hours per request and is at the upper end of what is practical in a design workflow. The same 22 target frequencies were used, giving 4{,}400 simulations. The study
required 253~hours ($\sim${10.5 days}) of continuous solver time.

The cost of the proposed method separates into two parts. The search evaluates
50{,}000 states through the surrogate and completes in 2.2~minutes, with no
solver calls. Verifying the ten candidates it returns takes 10 full-wave simulations, or
about 35~minutes, giving roughly half an hour per request. All timings were measured on
the same workstation, with an NVIDIA RTX 3090 used for both the full-wave
transient solver and the surrogate. Building the training dataset required 31{,}659 simulations, a cost incurred only
once. The solver-guided search needs 200 simulations for every request, so about 160
requests would cost as many simulations as the whole dataset.

\begin{figure}[!t]
\centering
\includegraphics[width=\columnwidth]{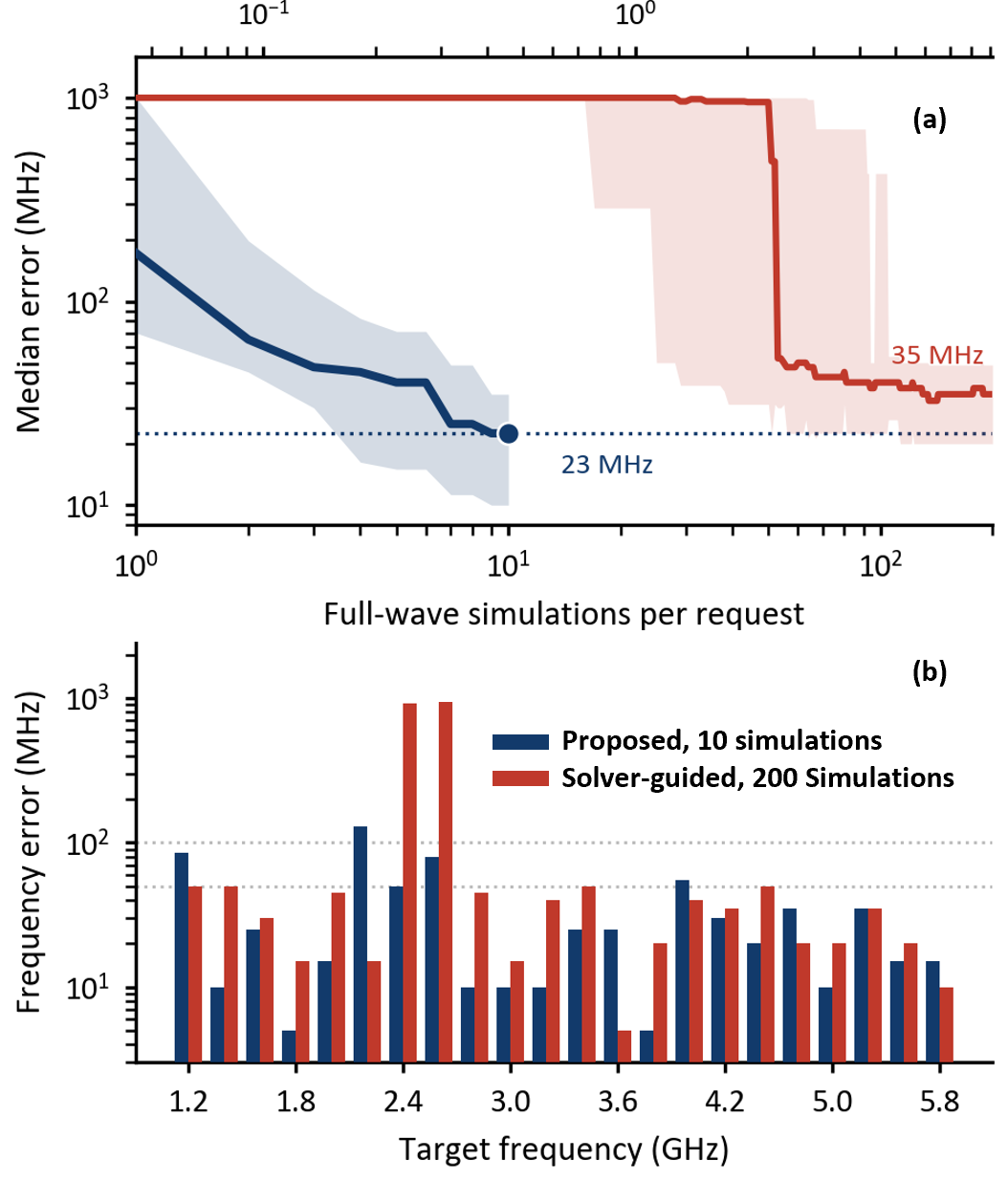}
\caption{Comparison with a solver-guided search over the same 22 targets.
(a) Median frequency error against the number of full-wave simulations spent per
request, with the equivalent solver time on the upper axis. Shaded bands are the
interquartile range. Where no resonance has yet been found inside the search
window the error is plotted at the 1~GHz window limit. (b) Best result for each
target frequency.}
\label{fig:baseline}
\end{figure}

Fig.~\ref{fig:baseline}(a) plots the median frequency error of both methods against
the number of simulations spent, with the corresponding solver time on the upper
axis. The proposed method reaches a median error of 23~MHz after its 10
simulations. The solver-guided search does not reach that level within its 200
simulations, ending at 35~MHz, so it uses twenty times the solver budget and
roughly twenty times the wall-clock time without matching the result. For the first
50 simulations it has typically found no resonance inside the target window at all,
because the objective is flat wherever no dip is present and the chain has no
gradient to follow. A graded probability is returned by the surrogate at every state, so
the search it drives always has a direction.

The two methods are compared target by target in Fig.~\ref{fig:baseline}(b). The proposed
method is closer at 15 of the 22 targets, with a median error of 23~MHz against
35~MHz and a mean of 32~MHz against 115~MHz. Every request is answered within 300~MHz, while the solver-guided search fails at
two: at 2.4~GHz a structure resonating at 1.48~GHz is returned and at 2.6~GHz one
resonating at 3.60~GHz, both well matched but at the wrong frequency.

The solver-guided search produces substantially deeper resonances, with a median depth of $-30.7$~dB compared with $-13.6$~dB for the proposed method. This difference is expected from the objectives used in the two searches. The objective in~(\ref{eq:objective}) maximises the predicted probability of a resonance occurring within the target bin, but does not directly optimise resonance depth. A shallow resonance at the correct frequency can therefore achieve a high objective value. The solver-guided search, in contrast, directly maximises the resonance depth and consequently favours deeper resonances.

% The comparison is not one-sided. Substantially deeper resonances are produced by the solver-guided search, with a median depth of $-30.7$~dB against $-13.6$~dB for the proposed
% method. This follows from the two objectives. The objective
% in~(\ref{eq:objective}) maximises the predicted probability that a resonance lies
% in the target bin and places no value on its depth, so a shallow resonance at the
% right frequency satisfies it. The solver-guided objective is the depth itself.

\section{Discussion}
\label{sec:discussion}
% =====================================================================

The forward surrogate reaches a mean per-bin AUC of 0.737, while the pipeline it
drives answers all 22 requests within 300~MHz through the best of their ten candidates. A moderate AUC is sufficient here because the surrogate is used only inside the
search, never as a stand-alone predictor. States are scored inside a search that is already restricted to valid patterns, so
only candidates that differ by one bridge have to be ordered, and the response of an
arbitrary pattern never has to be predicted. The chain is held inside the range of
pattern sizes present in the training set by the size penalty, so the states that are
scored stay close to the data the surrogate was fitted on. The final candidate performance is assessed through full-wave simulation, which is why the
method is reported against 220 simulated candidates rather than against the
surrogate's own test performance.

An extension of the method to two-port filter operation on the same board is reported in the remainder of this section, followed by the limitations of the present work.

\subsection{Extension to Two-Port Filter Operation}

A second mode of operation is supported by the same board. A second port, P2, sits at the
mirror of P1 on the opposite edge of the lattice, and is not used in the antenna work above. If the active bridges are made to join both ports rather than only P1,
the structure becomes a two-port network and the observable of interest changes from
the reflection coefficient alone to the pair $S_{\mathrm{11}}$ and $S_{\mathrm{21}}$. This was investigated as an extension of the present work, with
the aim of building a surrogate that could later be inverted to return a structure for
a requested passband. The attempt is reported here because its outcome bears on the
argument made in Section~\ref{sec:intro} about which kinds of response this board and
this modelling approach can support.

Patterns were produced by a generator built on the same principles as the antenna
generator of Section~\ref{sec:dataset}. Every pattern begins from a conducting path
that joins P1 to P2, either running straight along the port row or meandering within
one row of it, so a two-port connection exists by construction and no simulation is
spent on structures that cannot conduct. Additional metal is then placed around that
path from a fixed set of shapes: open and shorted stubs of bounded length, square and
concentric rings of bounded size, and vertical pillars held clear of the main line so
that they couple to it without shorting into it. The board is therefore not filled at
random. Each pattern is a two-port structure with a transmission path and a limited
number of loading elements. A total of 24{,}535 such patterns were simulated in CST
under the same protocol as the antenna dataset, with $S_{\mathrm{11}}$ and $S_{\mathrm{21}}$ recorded
from 1.0 to 6.0~GHz.

\subsubsection{What the Dataset Contains}
A filter must meet three conditions at once. It must pass the requested band with low
insertion loss, it must be matched over that band so the power enters the structure
rather than reflecting, and it must reject outside the band. Each pattern was
therefore tested for a contiguous region in which $S_{\mathrm{11}}$ falls below $-10$~dB and
$S_{\mathrm{21}}$ stays above $-3$~dB at the same frequencies, and then for rejection below
$-15$~dB in a 500~MHz window placed 100~MHz beyond each edge of that region.

The two conditions give very different counts. A passband satisfying the first
condition is present in 17{,}610 of the 24{,}535 patterns, or 71.8\%. When the rejection condition is added, 386 patterns are left, or 1.6\%. Passbands are therefore common on this board, whereas rejection beside them is rare. The rejection test is a local one, since it examines
only 500~MHz on either side of the passband rather than the whole sweep, and the board
still fails it in 98\% of cases.

\subsubsection{What the Surrogate Learned}
A forward surrogate was trained on the full set of 24{,}535 patterns to predict the
$S_{\mathrm{11}}$ and $S_{\mathrm{21}}$ curves directly, as a regression over the 1001 frequency points.
Unlike the antenna task, no reformulation was needed. The filter response varies
gradually across the band rather than concentrating in a few narrow dips, so a
regression objective is not dominated by a flat baseline and the difficulty described
in Section~\ref{sec:intro} does not arise in the same form.

\begin{figure}[!t]
\centering
\includegraphics[width=\columnwidth]{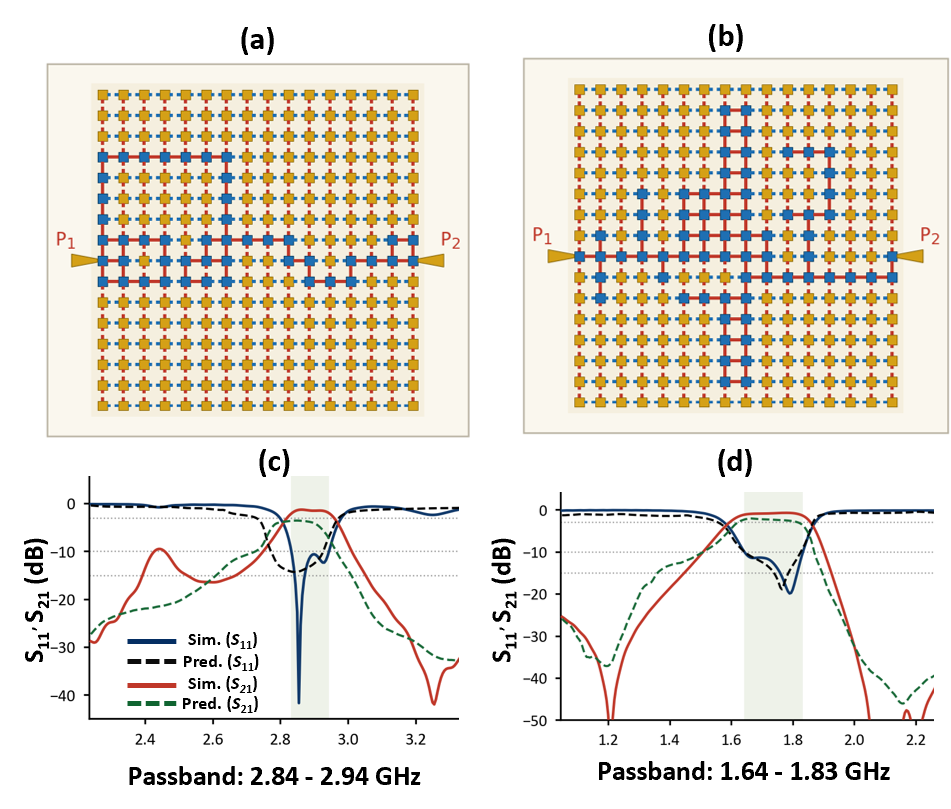}
\caption{Two-port filter operation on the same board. (a) and (b) Two patterns
with both ports connected. (c) and (d) The corresponding simulated $S_{\mathrm{11}}$ and $S_{\mathrm{21}}$, with the surrogate
prediction on the same axes and the shaded band marking the detected passband.}
\label{fig:filter}
\end{figure}

A mean absolute error of 1.85~dB on $S_{\mathrm{11}}$ and 4.02~dB on $S_{\mathrm{21}}$ is reached by the
trained model over a held-out test set. Two patterns and their responses are shown in
Fig.~\ref{fig:filter}, with the prediction plotted on the same axes. The main features
of both curves are placed at the right frequencies, including the transmission nulls
and the edges of the passband, but the predicted $S_{\mathrm{21}}$ level differs from the
simulation by up to 10~dB in places.

The $S_{\mathrm{21}}$ error is too large for the surrogate to drive an inverse search. A
passband requires $S_{\mathrm{21}}$ above $-3$~dB and a stopband requires it below $-15$~dB, so
the two are separated by 12~dB. A mean error of 4~dB is a third of that separation.
Transmission can therefore be reported as high or low in a given region, but a
passband cannot be reliably told apart from a stopband, which is the minimum requirement for a filter surrogate.

\subsubsection{Interpretation}
Two limits are visible, and both are present in the data before any question of
network architecture arises. The first is coverage. Only 1.6\% of the dataset contains a passband with adjacent rejection, so the target response is rarely represented in the training data. The second is
resolution. The broad form of $S_{\mathrm{11}}$ and $S_{\mathrm{21}}$ is learned by the surrogate, but not to the accuracy that separating a passband from a stopband requires.

The two modes impose different requirements. A resonance is a local feature. It occupies
a small part of the frequency axis, and a design method only has to place that one
feature correctly, which is what the classification formulation of
Section~\ref{sec:forward} does and what the full-wave verification of
Section~\ref{sec:results} confirms. A filter is a specification on the response across
the whole sweep, requiring transmission in one region and rejection in another at the
same time. On this board that outcome is rare, and it is not reached by the same
modelling approach.

\subsection{Limitations}
The proposed work with the antenna method has four limitations. First, every candidate is anchored at P1 by
the pattern generator of Section~\ref{sec:dataset}, which produces a training
distribution biased toward the left half of the lattice. Configurations that resonate
below 2~GHz often need to extend further across the grid, which the generator covers
poorly. A generator that grows from arbitrary seed locations would broaden the region
of the board the surrogate has seen and would likely improve performance at the low
end of the band.

Second, the bridges are modelled as ideal copper segments. In a real reconfigurable board, non-ideal switches such as PIN diodes or
radio-frequency micro-electro-mechanical systems are used, which add parasitic
resistance, capacitance, and inductance. These parasitics, together with the DC bias lines needed to control each switch, shift
the resonances and widen the bandwidth by amounts that depend on the switch
technology. Because the surrogate learns from simulated data, these effects can be
included by simulating the training patterns with a lumped switch model, with no
change to the classification formulation or the search. A
fabricated prototype with measured scattering parameters is required to establish the
gap between idealised simulation and hardware.

Third, the inverse search optimises for resonance frequency only. Radiation
pattern, polarization, and gain are not part of the objective and would require
additional surrogate outputs or a selection step applied to the candidate set.
Fourth, the presence of a resonance at the requested frequency is scored by the
objective of~(\ref{eq:objective}), but not its depth. The consequence is shown by the
comparison in Section~\ref{sec:results} where deeper resonances are produced by
a solver-guided search optimising depth directly. 
% Including a depth term in
% the objective would address this and requires no change to the surrogate.

% =====================================================================
\section{Conclusion}
\label{sec:conclusion}
% =====================================================================
In this paper, an inverse design method for a $16 \times 16$ reconfigurable pixel antenna with 480
binary switches has been presented and verified by full-wave simulation. The method
has two parts. The forward task is posed as multi-label classification over a
discretized frequency axis rather than as regression of the reflection-coefficient
curve, which suits a response whose useful content is a small number of narrow dips in
a wide sweep. The inverse task is a simulated annealing search over the switch state in
which the connectivity constraint is maintained at every step, so the chain only ever
visits patterns that are valid antennas, and a size penalty keeps it inside the region
of the input space the surrogate was trained on.

The method was tested end to end by generating candidates for 22 target
frequencies between 1.2 and 5.8~GHz and simulating all 220 of them in CST. All 22 requests were answered within 300~MHz by the best of their ten candidates, and 80\% of the detected resonances reached below $-10$~dB. A mean per-bin AUC of 0.737 on a held-out test set is reached by the forward
surrogate that guides the search, which is sufficient to rank the valid candidate patterns evaluated during the search and identify configurations whose resonant frequencies approach the requested target. A search that scores every candidate with the full-wave solver instead of the
surrogate reaches a median error of 35~MHz after 200 simulations per request,
against 23~MHz after 10 for the method presented here.

The same board was also examined in two-port operation, where the objective is a filter
response rather than a resonance. That extension did not reach the same standard.
Passbands are common in the simulated data, appearing in 71.8\% of patterns, but
only 1.6\% also reject the frequencies immediately beside the passband, and a
surrogate trained on the full set predicts $S_{\mathrm{21}}$ with a mean absolute error of 4.02~dB, which is too
coarse to separate a passband from a stopband. Placing one narrow feature at a
requested frequency is a task this board and this method perform well, while specifying
the response across the whole sweep is not.

Three directions follow from this work. The first concerns coverage of the board.
Every training pattern is anchored at P1 by the pattern generator, so extending it to
grow from arbitrary seed locations would broaden the region of the board the surrogate
has seen and should improve performance at the low end of the band, where large
connected regions are needed. The second concerns hardware. The switches are modelled
as ideal copper, so a fabricated prototype with measured scattering parameters is
required to establish the gap between simulation and hardware. Such a board would
also allow the training data to be collected by measurement, since each switch state
can be set electronically and its reflection coefficient recorded with a vector
network analyser in seconds. A surrogate trained on measured data would then include
the effects of real switches and bias lines directly. The third is filter operation.
A much larger dataset weighted toward topologies that produce rejection would be
needed, together with a training objective that targets the passband and stopband
levels directly rather than the mean error across the curve, before the same approach
could be applied.
% Bibliography
% =====================================================================
\bibliographystyle{IEEEtran}
\bibliography{references}
\vspace{-2mm}
% =====================================================================
% Author biographies
% Replace each placeholder image with the corresponding author's photo.
% Replace the bracketed/sample biography text with verified author details.
% =====================================================================
\begin{IEEEbiography}
[{\includegraphics[width=1in,height=1.25in,clip,keepaspectratio]{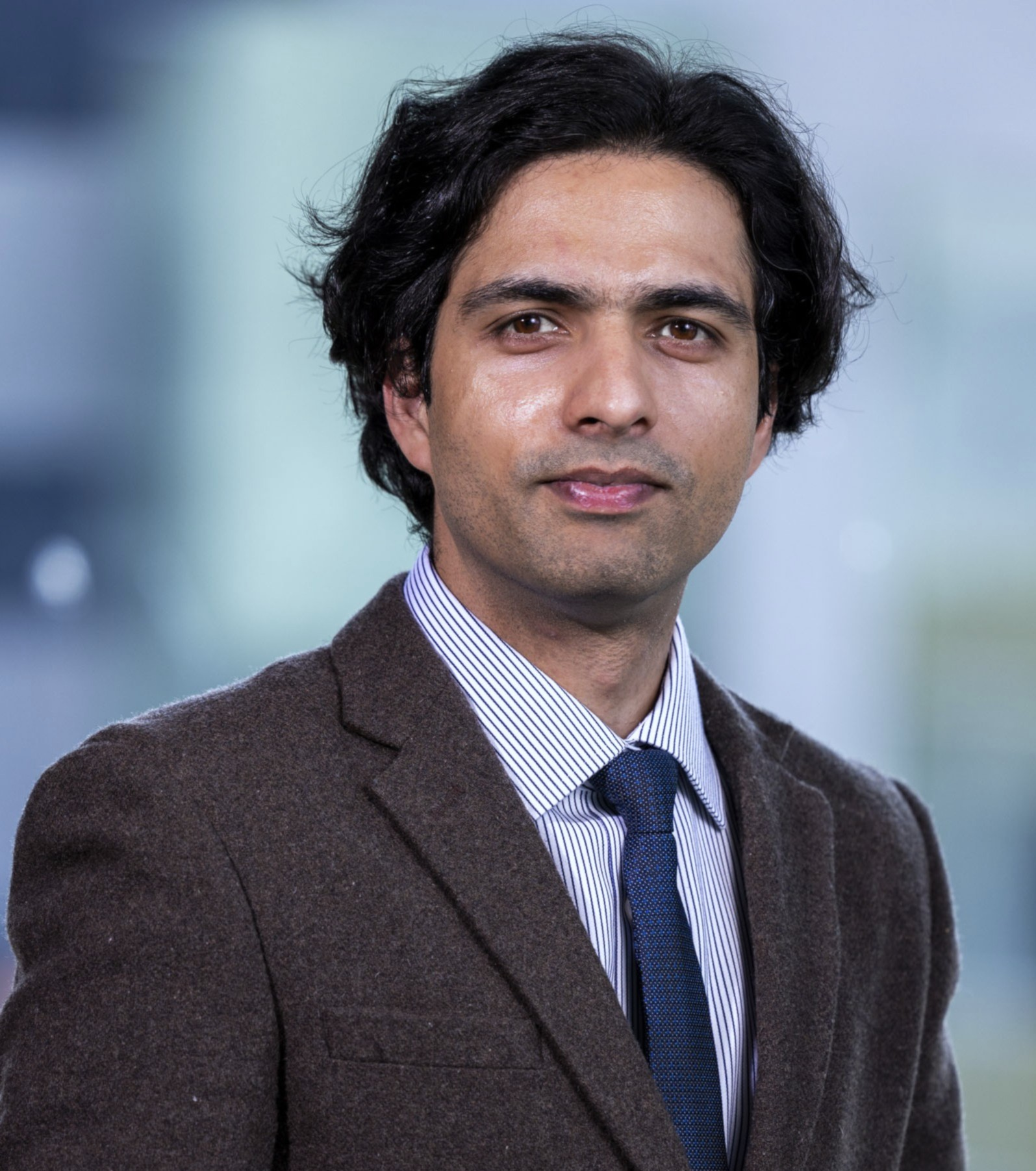}}]
{Nadeem Rather} is a Research Engineer with the Wireless Communications and AI Research Group at Tyndall National Institute, University College Cork (UCC), Ireland. He received the Ph.D. degree in electronic engineering from UCC in 2024. His research focuses on artificial intelligence (AI)-enabled wireless systems, chipless radio-frequency identification (RFID) sensing, electromagnetic system optimisation, and advanced antenna technologies. His research interests also include deep learning for radar-based sensing, reconfigurable intelligent surfaces, and intelligent electromagnetic structures for next-generation Internet of Things applications. He has received several research distinctions, including the Wrixon Research Excellence Bursary in 2024, the Best Application of AI in a Student Project award at the AI Awards Ireland 2024, and the Best Student Paper Award at the International Workshop on Antenna Technology. He also received the Best Technology Demonstration Award at the VistaMilk Internal Conference in 2023 and was recognised as a runner-up in Tyndall National Institute’s Postgraduate Publication of the Year competition in 2023. 
\end{IEEEbiography}
\vspace{-5mm}
\begin{IEEEbiography}
[{\includegraphics[width=1in,height=1.25in,clip,keepaspectratio]{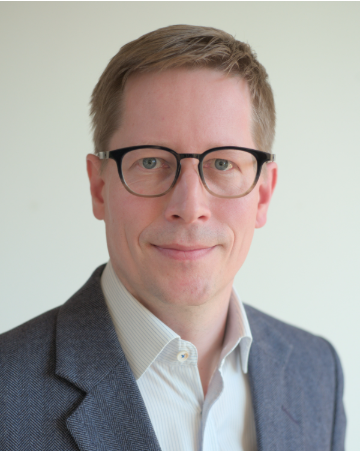}}]
{Holger Claussen}
Prof. Holger Claussen is Head of the Wireless Communications Laboratory at Tyndall National Institute, and Professor of Wireless Communications at University College Cork and Trinity College Dublin. At Tyndall, he is building up research teams in RF, Access, Protocols, AI, and Quantum Systems to invent the future of Wireless Communications. Previously, he led the Wireless Communications Research Department of Nokia Bell Labs in Ireland and the US until 2020. He received the 2014 World Technology Award in Communications Technologies for innovative work of ``the greatest likely long-term significance'' for his contributions to Small Cell Networks. Holger received his Ph.D. degree from the University of Edinburgh, UK in 2004. He is author of the book ``Small Cell Networks'', more than 140 journal and conference publications, 87 granted patent families, and 39 filed patent applications pending. He is Member of the Royal Irish Academy, Fellow of the IEEE, and Fellow of Trinity College Dublin.
\end{IEEEbiography}
\vspace{-10mm}
\begin{IEEEbiography}
[{\includegraphics[width=1in,height=1.25in,clip,keepaspectratio]{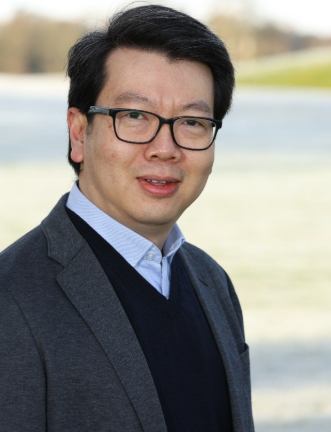}}]
{Lester Ho}
is a Principal Scientist at the Tyndall National Institute in Dublin where he leads the Future Access team. Previously, he spent 20 years performing research in wireless networks in Nokia Bell Labs, where he was a Distinguished Member of Technical Staff. There, he performed research in the areas of self-optimizing networks, cellular architectures, and network energy efficiency. He has invented the self-optimization techniques and scalable architectures that enabled the commercial deployment of small cells, where many of his innovations are used in live cellular networks today. His current research interests are in network optimization at the physical to network layers, O-RAN, machine learning, quantum repeater networks, and wireless network simulations.
\end{IEEEbiography}

\EOD
\end{document}